\documentclass[%
 reprint,
 amsmath,amssymb,
 aps,
]{revtex4-2}

\usepackage{graphicx}
\usepackage{dcolumn}
\usepackage{bm}

\begin{document}

\preprint{APS/123-QED}

\title{Tunneling control of dipolar boson in triple well circuit via dipole polarization orientation and quantum sensing application}



\author{Leandro H. Ymai}
\author{Arlei P. Tonel}
\email{arleitonel@unipampa.edu.br}
\affiliation{ Universidade Federal do Pampa, \\
Travessa 45, n 1650, \\
Bairro Malafaia, Bag\'e, RS, Brazil}%


\begin{abstract}
Precise control of the transport of ultra-cold atoms in optical lattices is essential for the exploration of quantum phenomena and the development of advanced quantum technologies. Harnessing the distinctive characteristics of the system near integrability, we investigate the tunneling control of dipolar bosons confined within a triple-well circuit via dipole polarization orientation. Based on this control mechanism, we propose a quantum magnetic compass that shows sensitivity beyond the Heisenberg limit to small variations in the orientation of magnetic fields, making it a promising candidate for sensing applications.
\end{abstract}

\maketitle


\section{\label{sec:level1}Introduction}
The development of technologies based on quantum principles can lead to innovations with a potentially transformative impact in various fields. Among the platforms for exploring and harnessing these quantum properties, such as superconducting circuits \cite{Xiang2013, Christensen2020}, integrated quantum photonics \cite{Wang2020} and spin chain \cite{Marchukov2016}, optical lattices with the capacity to confine ultracold atoms in crystalline arrangements of potential wells offer a versatile alternative for this purpose \cite{RevModPhys.80.885}. Recent progress in quantum gas trapping has made it possible to isolate small lattice cells in various geometries \cite{Windpassinger_2013} with high level of control of system parameters during an experimental sequence, providing ideal testbeds for proposing novel quantum devices with prospect of applications in quantum simulation, quantum computing and quantum sensing  \cite{Amico2017, PhysRevLett.94.090405, RevModPhys.89.035002}. From a theoretical perspective, the behavior of atoms within these confined lattice cells can be described using tractable physical models,
providing insights into fundamental physics and guide future experiments.
The Extended Bose-Hubbard (EBH) model \cite{PhysRevB.40.546,PhysRevB.47.279,Baier201,Lahaye_2009,PhysRevA.73.013625}  is a fundamental example of physical model that has been extensively studied in condensed matter physics and incorporates additional inter-site dipole-dipole interaction (DDI) in addition to on-site short-range interactions and tunnelling between sites to provide an adequate description of dipolar bosons in optical lattices. The long-range and anisotropic nature of DDI is an additional ingredient that makes it possible to explore richer regimes and scenarios that favour the emergence of quantum phenomena \cite{Goral2002, Menotti2007}.

Typically, models derived from the EBH for a few  sites are non-integrable. However, precise control over experimental parameters makes it possible to fine-tune them, yielding integrable \cite{RevModPhys.73.307,JonLinks_2003,Tonel_2015,Ymai_2017} or even superintegrable models \cite{BENNETT2024116406}, providing a rigorous theoretical framework for addressing questions that can not be answered analytically using non-integrable theories. The quantum inverse scattering method and the algebraic Bethe ansatz \cite{korepin1993quantuminversescatteringmethod,Sklyanin:1979pfu,10.1007/3-540-11190-5_8,Faddeev1994AlgebraicAO} have been powerful tools for building integrable models to study bosonic atoms in lattices of few wells \cite{JonLinks_2003,  Tonel_2015,Ymai_2017} and device proposals such as transistors \cite{PhysRevA.75.013608,PhysRevA.75.023615, Wilsmann_2018} and interferometers \cite{PhysRevLett.104.170404,PhysRevLett.129.020401}.

In this work, we investigate a system of dipolar atoms confined in a triple-well ring structure. We derive the Hamiltonian from the EBH model in the vicinity of the specific direction of dipole polarization in which the Hamiltonian is integrable. Next, in order to obtain a reliable regime for the optimal operation of the system, we analyze the energy spectrum of the integrable Hamiltonian and identify the range of parameters values and corresponding initial conditions in which the atomic populations exhibit robust resonant behavior in response to the zenith angle variation of the dipole polarization direction. This analysis allows us to derive an effective Hamiltonian from which we can obtain analytical expressions for relevant observables, making it possible to investigate the control of tunnelling through the orientation of the dipole polarization. Using our findings, we outline a protocol to implement a quantum magnetic compass and show its high sensitivity to the variation in direction of a uniform magnetic field acting on the system, surpassing the Heisenberg limit.

The paper is organized as follows: Section \ref{secII} presents the theoretical framework and the main parameters of the system based on the EBH model. Section \ref{secIII} focuses on the integrable Hamiltonian and the scenarios that characterise a robust resonant regime of atomic populations. In Section \ref{secIV}, we analyse the dynamics of the system under small changes in the polarization direction near integrability. Section \ref{secV} presents the protocol for a quantum magnetic compass and analyses its sensitivity. Finally, Section \ref{secVI} concludes the article with a discussion of the results and potential future directions.

\section{\label{secII}Theoretical framework}
In this section, we present an overview of the system, analyzing its main features that will be used throughout the text. We consider a system of ultracold dipolar atoms (e.g., dysprosium or erbium) confined into triple-well potential in ring configuration. The many-body physics description of this system can be captured by the extended Bose-Hubbard model \cite{Baranov2008}, given by the Hamiltonian

\begin{eqnarray}
H &=& \frac{U_0}{2}\sum_{i=1}^3N_i(N_i-1)+\sum_{i,j=1}^3\frac{U_{ij}}{2}N_iN_j \nonumber \\ &+&\sum_{i=1}^3\nu_iN_i-J\sum_{i<j}^3(a_i^\dagger a_j + a_j^\dagger a_i),\nonumber
\end{eqnarray}
where $a_i^\dagger$ ($a_i$) is the boson creation (annihilation) operator, $N_i$ represents the boson occupation at the site $i$ with total number of particles $N=N_1+N_2+N_3$ conserved, $U_0$ is the on-site interaction energy resulting from the contact interaction and DDI between particles at the same site, $U_{ij}=U_{ji}$ is the interaction energy between particles at site $i$ and $j$ due the DDI, $\nu_i$ represents the energy gradient of site $i$ resulting from local external field and $J$ is the hopping rate between the sites. The dipoles can be polarized in a given direction by a magnetic field and the long-range interaction energy is given by
\begin{eqnarray}
U_{ij} = \widetilde{U}(1-3\cos^2\theta_{ij}),\nonumber
\end{eqnarray}
where $\theta_{ij}$ is the angle between the polarization direction and the relative position of sites $i$ and $j$ and $\widetilde{U}$ is the interaction energy when $\theta_{ij} = \pi/2$.  
For a general direction of dipole polarization, represented by the unit vector (see Figure 1):
\begin{eqnarray}
{\bf d} = (\sin\phi \sin \theta,\sin\phi\cos\theta,\cos\phi),\nonumber
\end{eqnarray}
where $\theta$ and $\phi$ are the polar and zenith angles, respectively, the interaction energy can be calculated by the formula
\begin{eqnarray}
U_{ij}=\widetilde{U}[1-3\,({\bf d}\cdot {\bf u}_{ij})^2],\nonumber
\end{eqnarray}
where ${\bf u}_{ij}$ is the unit vector of relative position between the sites $i$ and $j$.
\begin{figure*}[!ht]
\center
\includegraphics[scale = 0.7]{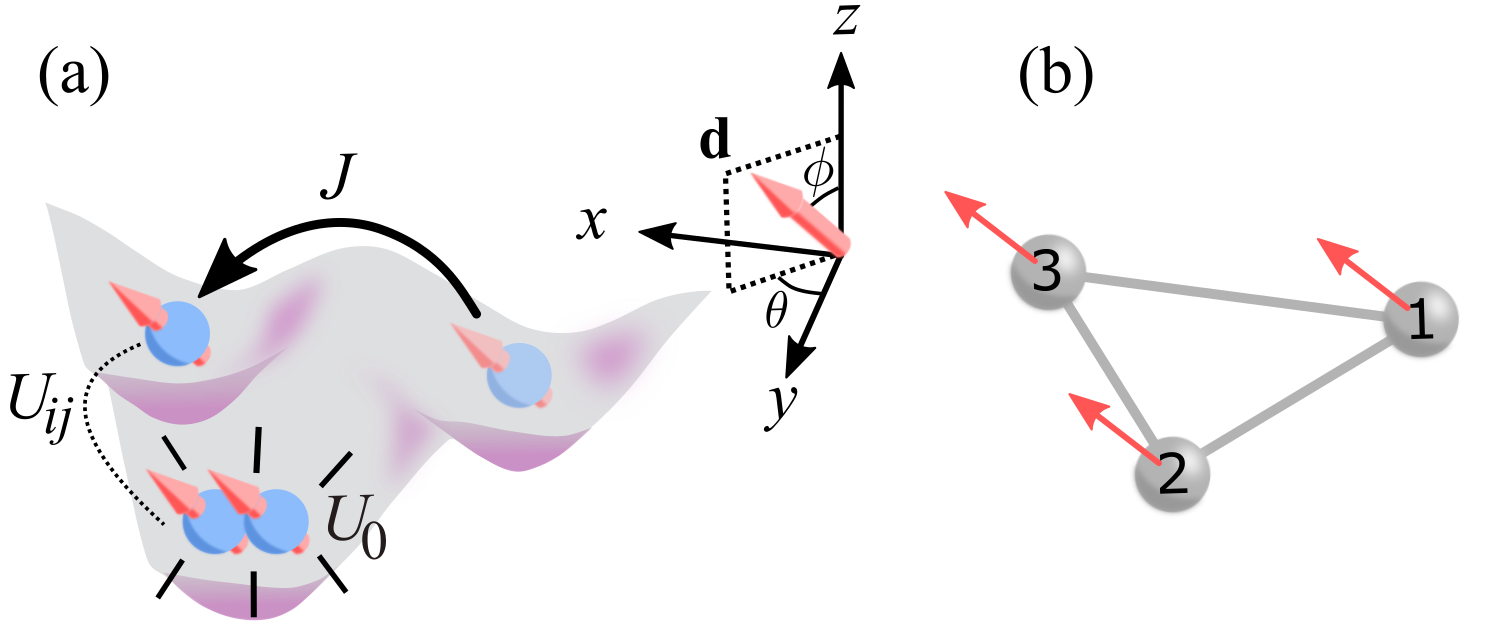}
\caption{(a) Triple well potential in ring configuration. The blue spheres represent the dipolar atoms and the red arrows represent the direction of dipole polarization characterized by the polar and zenith angles $\theta$ and $\phi$, respectively. The parameters $U_0$ and $U_{ij}$ represent the on-site and inter-site interaction energies. Particles can tunneling through the sites with hopping rate $J$. (b) Schematic graph representation of the system.}
\end{figure*}
Our focus is on the case of spherical potential wells in which the on-site energy results only from the contact interaction \cite{Lahaye_2009}, and can be balanced with the long-range interaction energy when all dipoles are oriented along $z$-direction, satisfying
\begin{eqnarray}
U_0 = \widetilde{U}.
\end{eqnarray}
This condition can be achieved by tuning the depth of the potential wells and the scattering length. The details and tolerances of this condition are given in Appendix \ref{appC}.  
In this case, the Hamiltonian can be conveniently rewritten (see Appendix \ref{appA} for details) as
\begin{widetext}
\begin{eqnarray}
H &=& \frac{U}{3}[2+\cos(2\theta)](N_1-N_2+N_3)^2-\frac{16}{3}U\left[\sin^2\theta N_1N_3+\frac{\sqrt{3}}{4}\sin(2\theta)N_2(N_1-N_3)\right] 
\nonumber \\ &+&\nu(N_1-N_2+N_3)- J\sum_{i<j}^3(a_i^\dagger a_j+a_j^\dagger a_i),\label{Hg}
\end{eqnarray}
\end{widetext}
where we define
\begin{eqnarray}
U = \frac{9}{16}\widetilde{U}\sin^2\phi,\nonumber
\end{eqnarray}
and $\nu$ is energy gradient between the site 2 and the subsystem of sites 1 and 3, which can be implemented by a harmonic potential with the center slightly displaced toward site 2 (see more details in Appendix \ref{appB}).  

The emphasis of our work in what follows is to analyze the behavior of the atomic flow throughout the sites under small variations in the orientation of the dipole polarization in the vicinity of $\theta = 0$. Thus, the first order expansion of the Hamiltonian in $\theta$ parameter leads to the Hamiltonian
\begin{eqnarray}
H &=& U(N_1-N_2+N_3)^2+\gamma\theta N_2(N_1-N_3) \nonumber\\
&&+\nu(N_1-N_2+N_3)- J\sum_{i<j}^3(a_i^\dagger a_j+a_j^\dagger a_i),\label{H}
\end{eqnarray}
where we define
\begin{eqnarray}
\gamma = -\frac{8}{\sqrt{3}}U.\nonumber
\end{eqnarray}
In the next section, we will analyze the range of parameters that leads to the dynamics of interest.

\section{\label{secIII}Integrable structure and resonant regime}
\begin{figure}[!htb]
\center
\includegraphics[scale=0.40
]{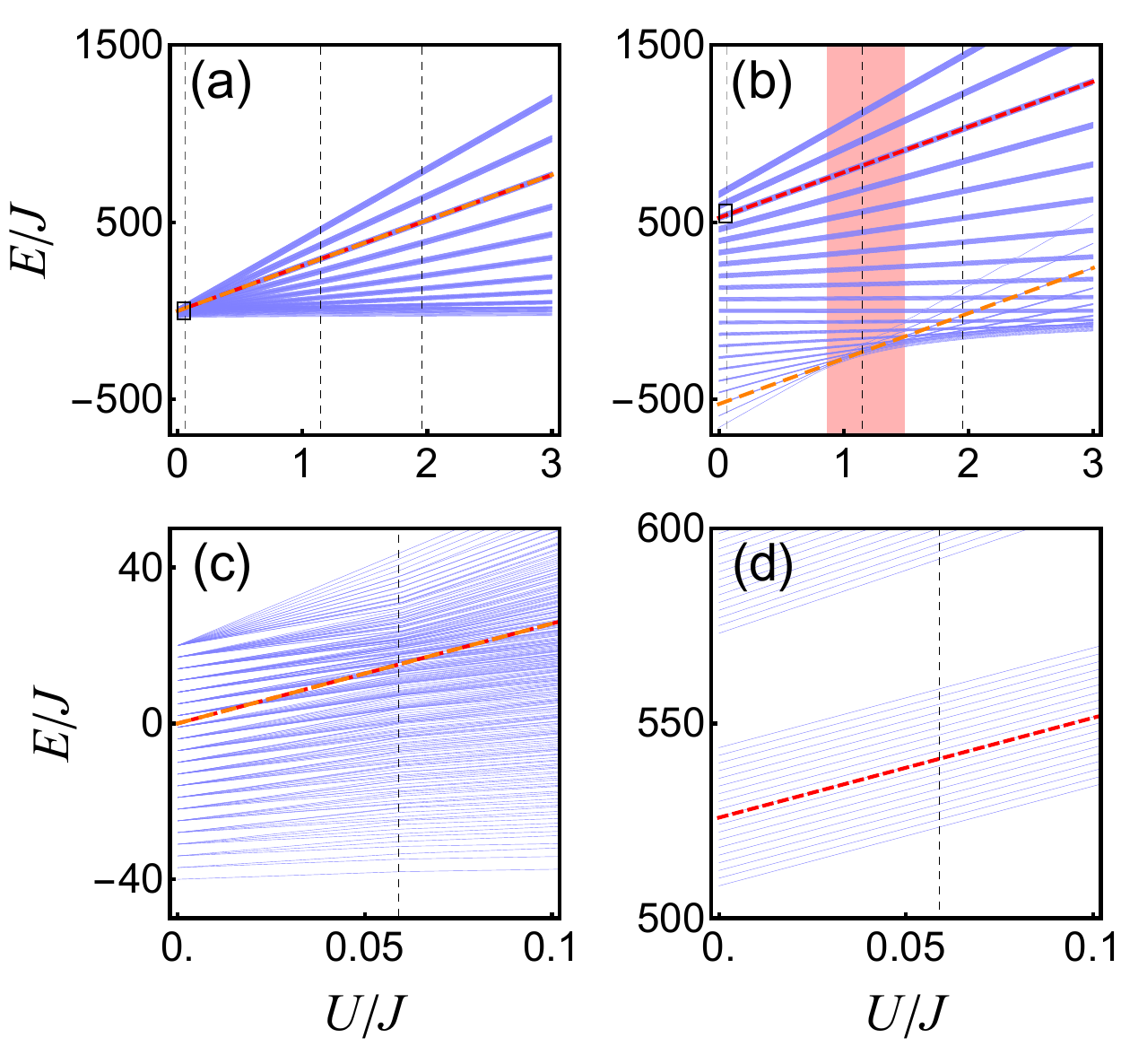}
\caption{$E/J$ versus $U/J$. (a) $\nu/h =0$ and (b) $\nu/h = 50.94$ Hz. Figures (c) and (d) show the regions highlighted by the rectangles in figures (a) and (b) enlarged. The vertical dashed lines represent the ratio $U/J$ for $\phi = $ 10, 50 and 90 degrees. The colour dashed lines represent the eigenvalues $E_{18,2,0}/J$ (red) and $E_{2,18,0}/J$ (orange).}
\label{esp}
\end{figure}

In this section, we analyze the mathematical structure of the Hamiltonian \eqref{H} and discuss the range of parameter values in order to obtain dynamic regimes characterized by the resonant tunneling behavior of atomic populations between sites.
For the case $\theta = 0$, the Hamiltonian \eqref{H} can be decomposed in three parts
\begin{eqnarray}
H = H_{\rm int} +2JQ-\frac{JN}{2},\label{Hint}
\end{eqnarray}
where
\begin{eqnarray}
H_{\rm int} &=& U(N_1-N_2+N_3)^2+\sigma(N_1-N_2+N_3)\nonumber \\ &-&  J[a_2^\dagger(a_1+a_3)+(a_1^\dagger+a_3^\dagger)a_2],\nonumber
\end{eqnarray}
is an integrable Hamiltonian  \cite{Ymai_2017}, $\sigma = \nu-J/2$ and 
\begin{eqnarray}
Q = \frac{1}{2}(N_1+N_3-a_1^\dagger a_3-a_3^\dagger a_1),\nonumber
\end{eqnarray}
is a conserved operator. The last term is a global constant and as $[H_{\rm int}, Q] = 0$, the Hamiltonian \eqref{Hint} is also integrable. The elucidation of this integrable structure of the Hamiltonian allows us to analyze the global behavior of the system through the specific functionalities of its subsystems in a regime characterized by the formation of bands in the system's energy spectrum. To examine the parameter values that result in a band structure in the energy spectrum and its implications for the system's dynamics, we first note that the eigenstates of the Hamiltonian  \eqref{Hint} at $J = 0$ are given by the Fock states:
\begin{small}
\begin{eqnarray}
|n_1,n_2,n_3\rangle =\frac{(a_1^\dagger)^{n_1}}{\sqrt{n_1!}}\frac{(a_2^\dagger)^{n_2}}{\sqrt{n_2!}}\frac{(a_3^\dagger)^{n_3}}{\sqrt{n_3!}}|0\rangle, \quad n_1+n_2+n_3 = N,\nonumber
\end{eqnarray}
\end{small}
with corresponding eigenvalues $E_{n_1,n_2,n_3} = U(n_1-n_2+n_3)^2+\nu (n_1-n_2+n_3)$. Then, when the interaction and external field terms of the Hamiltonian dominate over the tunneling term, we observe the formation of bands of approximate width $|2(N-n_2+1)J|$ enclosing the eigenvalues $E_{n_1,n_2,n_3}$, as can be observed from the energy spectrum of the Hamiltonian \eqref{Hint} in Figure \ref{esp}. For illustrative purposes, we will proceed with the parameter values $U/h = 5.38$ Hz and $J/h = 1.55$ Hz, which are specified in Appendix \ref{appC} for $N = 20$ bosons. However, we emphasize that these are not the only valid parameters and other options are available.  
It can be seen from Figure \ref{esp}-(a) that for the case of $\nu = 0$ the energy spectrum shows the formation of bands as $U/J$ increases. Figure \ref{esp}-(c) shows the rectangular region of Figure \ref{esp}-(a) enlarged, with the energy levels accumulating without the band structure for small values of $U/J$. On the other hand, Figure \ref{esp}-(b) shows the case of parameter $\nu$ large enough for the band structure to be restored in the region of small values of $U/J$. In order to quantify the value of $\nu$ needed to obtain the band separation, we recall that the widths of the highest bands are $2NJ$ and $2(N+1)J$. This allows us to estimate the separation between the highest band as
\begin{eqnarray}
\Delta E = [E_{N,0,0}|_{U=0}-(N+1)J] -[E_{N-1,1,0}|_{U=0}+NJ].\nonumber 
\end{eqnarray}
Thus, the completely separation of the bands in the region of small values of $U/J$ occur when $\Delta E > 0$, which leads to the condition $\nu > (N+1/2)J$. Figure \ref{esp}-(d) shows the rectangular region of Figure \ref{esp}-(b) enlarged, indicating that the energy levels within the band are uniformly distributed, similar to the energy levels of a harmonic oscillator. This aspect causes the expectation values of the populations to exhibit resonant behavior, as we will see later. Hence, in the regime in which the bands are formed, we call it a {\it resonant regime}, and the dynamics of atomic populations can be described by the expectation values $\langle N_i \rangle \equiv \langle \Psi(t)|N_i|\Psi(t)\rangle$, where the quantum state of the system is given by
\begin{eqnarray}
|\Psi(t)\rangle = \exp(-it H)|\Psi(0)\rangle,\nonumber
\end{eqnarray}
(we set $\hbar =1$). In resonant regime the mainly contribution to dynamics result from eigenstates of the energy levels within the band determined by the initial Fock state $|\Psi(0)\rangle = |n_1,n_2,n_3\rangle$.  The predominance of the interaction and the external potential terms over the tunneling term leads to second-order processes that allows us to derive an effective Hamiltonian dependent of conserved operator \cite{Wilsmann_2018}, given by
\begin{eqnarray}
H_{\rm eff} = \mathcal{J}Q,\nonumber
\label{Heff}
\end{eqnarray}
where the effective hopping parameter is given by
\begin{small}
\begin{eqnarray}
\mathcal{J} = 2J-\frac{J^2}{2U}\left(\frac{n_2+1}{N-2n_2-1+\frac{\sigma}{2U}}-\frac{n_2}{N-2n_2+1+\frac{\sigma}{2U}}\right).\nonumber \\\label{Heff1}
\end{eqnarray}
\end{small}
It is clear that the effective Hamiltonian is independent of the mode of site 2, which implies that the dynamics occur in the subsystem of sites 1 and 3, maintaining the expectation values of site 2 constant, i.e., $\langle N_2\rangle = n_2$. The effective Hamiltonian allows us to obtain explicit formulas for the expectation values of populations of sites 1 and 3, given by
\begin{eqnarray}
\langle N_1\rangle &=& n_1+(n_3-n_1)\sin^2\left(\mathcal{J}t/2\right),\label{nk1}\\
\langle N_3\rangle &=& n_3-(n_3-n_1)\sin^2\left(\mathcal{J}t/2\right),\nonumber
\end{eqnarray}
which clearly shows that $\langle N_1+N_3\rangle = n_1+n_3$ is also conserved. 

\begin{figure}[!htb]
\center
\includegraphics[scale=0.35]{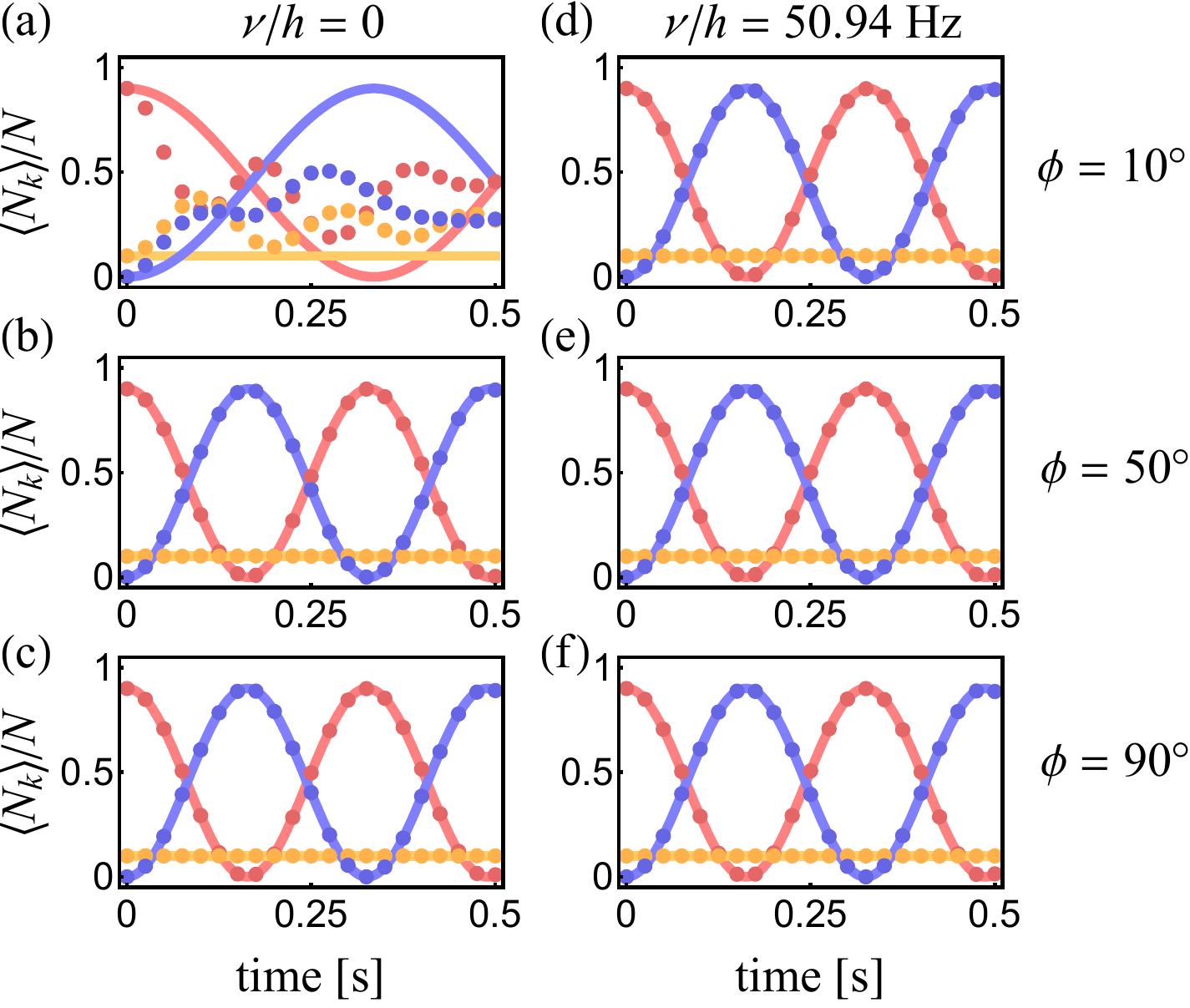}
\caption{Time evolution of fractional population $\langle N_k \rangle/N$ for $k=1$ (red), $k=2$ (orange) and $k=3$ (blue). The dotted curves result from the numerical simulation using Hamiltonian \eqref{Hint} and the solid curves represent the theoretical prediction \eqref{nk1}. We consider the initial state $|\Psi(0)\rangle = |18,2,0\rangle$ for $\nu/h = 0$ (left column) and $\nu/h = 50.94$ Hz (right column). From top to bottom, we consider interaction energy $U=\frac{9}{16}\widetilde{U}\sin^2\phi$ with respective parameters $\phi =$ 10, 50 and 90 degrees.}
\label{dina1}
\end{figure}

\begin{figure}[!htb]
\center
\includegraphics[scale=0.36]{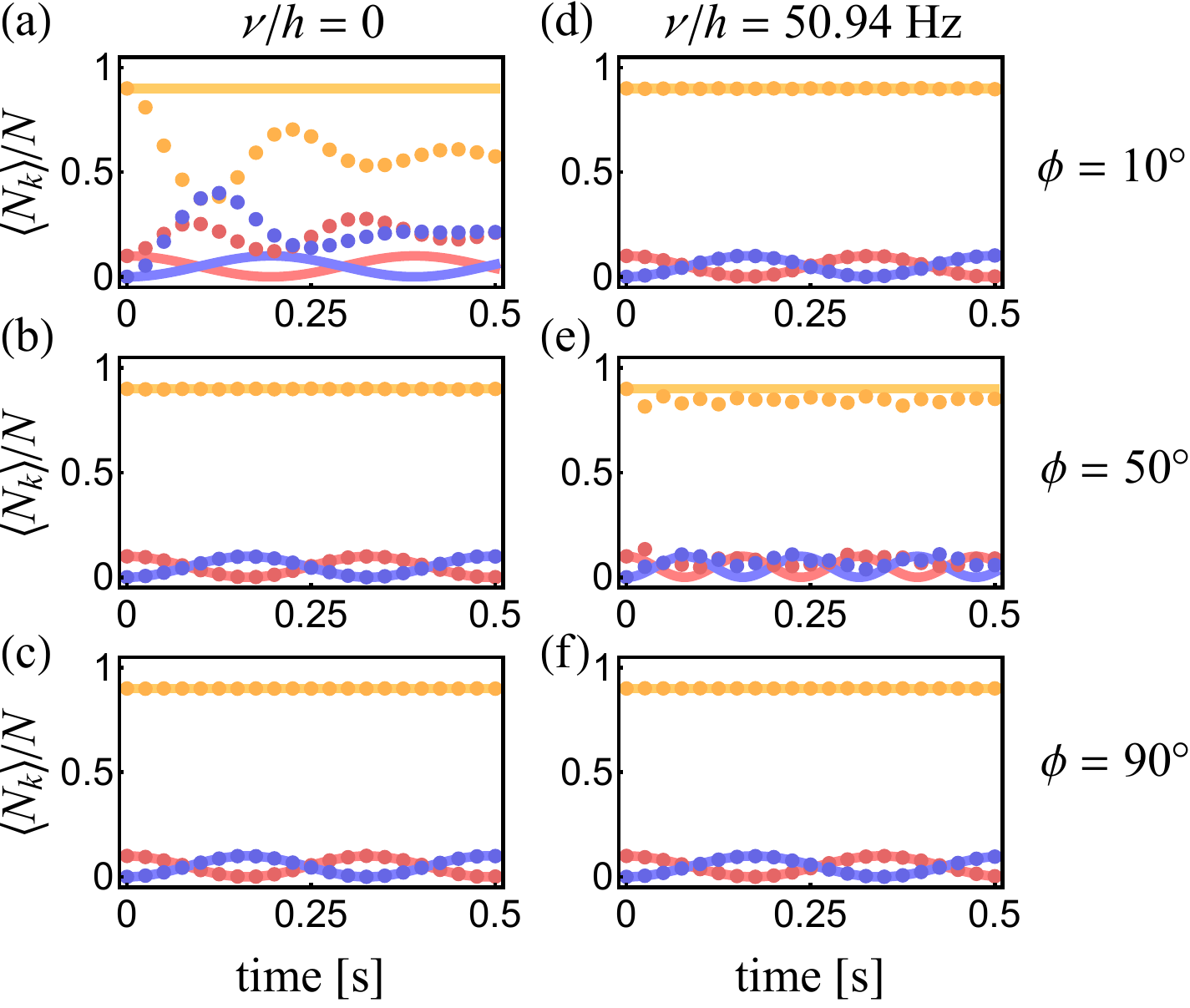}
\caption{The parameters are the same as in Figure \ref{dina1} but now using the initial state $|\Psi(0)\rangle = |2,18,0\rangle $.}
\label{dina2}
\end{figure}

Next, we examine two opposite extreme initial conditions, with $n_1+n_3\gg n_2$ and $n_1+n_3\ll n_2$, which present different aspects with respect to the structure of band formation in the case $\nu > (N+1/2)J$, as can be seen from Figure \ref{esp}-(b). In the first case $n_1+n_3\gg n_2$, the band remains formed in a broad domain of $U/J$ values, while in case $n_1+n_3\ll n_2$, there is an intermediate region $\nu/[J(2N-2)]<U/J<\nu/[J(N+2)]$ of band coalescence (see the pink region in Figure \ref{esp}-(b)).  Figures \ref{dina1} and \ref{dina2} compare the dynamics with $n_1+n_3\gg n_2$ and $n_1+n_3\ll n_2$, respectively, for cases $\nu = 0$ and $\nu>(N+1/2)J$, using the angle values $\phi$ corresponding to the dashed vertical lines in Figure \eqref{esp}. The left column of Figures \eqref{dina1} and \eqref{dina2}, shows disagreement between numerical results and formulas \eqref{nk1} in the first row (Figures \eqref{dina1}-a and \eqref{dina2}-a) where the system is out of the resonant regime, due to the nonexistence of band formation, as in Figure \eqref{esp}-(c).
In contrast, the right column of Figure \eqref{dina1} shows total agreement between the numerical result and the formulas \eqref{nk1} for a wide range values of parameter $\phi$, indicating the robustness against variations of the zenith angle $\phi$ at $\theta = 0$ for the case of initial state with $n_1+n_3\gg n_2$. The right column of Figure \eqref{dina2} shows disagreement between numerical results and formulas \eqref{nk1} in the second row (Figure \eqref{dina2}-(e)), due to the region of band coalescence in Figure \eqref{esp}-(b). 
The above analysis shows that, for the case $n_1+n_3\gg n_2$, the condition $\nu>(N+1/2)J$  is a more reliable option to maintain a wide range of variation of the $\phi$ parameter in the resonant regime.  On the other hand, for the case $n_1+n_3\ll n_2$, the condition $\nu = 0$ ensures a longer uninterrupted interval of gradual variation of the $\phi$ parameter that maintains the resonant regime. This result will be useful in the analysis of the effect of a small variation in the orientation of the polarization of the dipoles in the vicinity of $\theta = 0$ on the dynamics, to be discussed in the following section.

\section{\label{secIV}Tunneling control via dipole polarization orientation}
\begin{figure}[h]
\center
\includegraphics[scale=0.35]{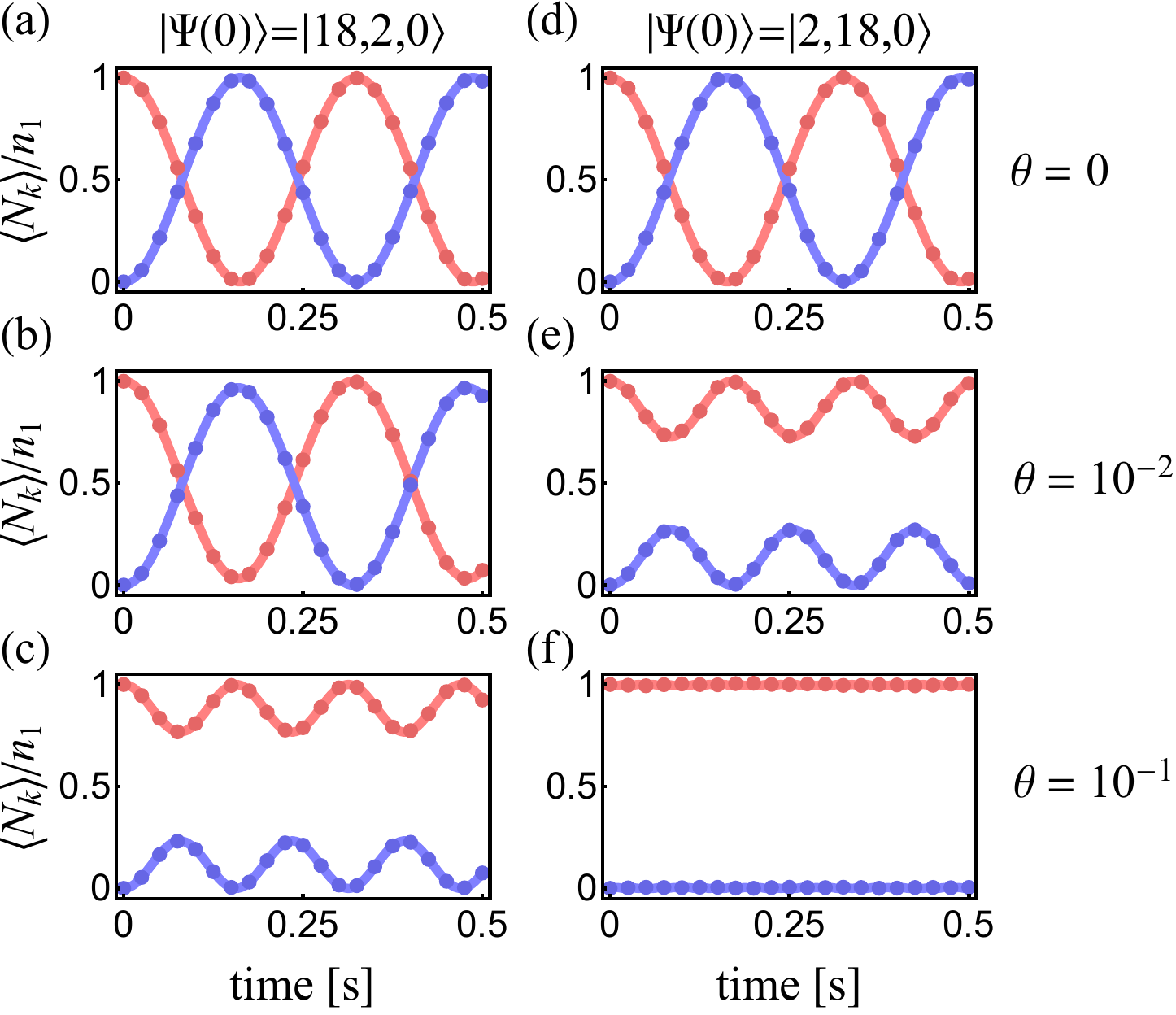}
\caption{Time evolution of $\langle N_k \rangle/n_1$. $k=1$ (blue) and $k=3$ (red).  We consider $\nu/h = 50.94$ Hz for initial state $|\Psi(0)\rangle =|18,2,0\rangle$ (left column) and $\nu/h = 0$ for $|\Psi(0)\rangle =|2,18,0\rangle$ (right column). For the interaction energy $U=\frac{9}{16}\widetilde{U}\sin^2\phi$ we consider $\phi = 90^\circ$, for the respective values from top to bottom: $\theta = $ 0, $10^{-2}$,  $10^{-1}$ rad. The solid lines represent the results obtained from \eqref{nk2} and dotted curves result from numerical simulation using the Hamiltonian \eqref{H}.}
\label{dtheta}
\end{figure}

Now we consider the effect of a small change in the direction of dipole polarization in the vicinity of $\theta = 0$ on the tunneling process. Since expectation value at site 2 remains constant as $\langle N_2\rangle = n_2$ in resonant regime, the effective Hamiltonian can be written as
\begin{eqnarray}
H_{\rm eff} =\mathcal{J}Q+\gamma\,\theta\,n_2 (N_1-N_3). \nonumber
\end{eqnarray}
Next, by ignoring the constant $N_1+N_3$ and make the change of basis of bosonic operators
\begin{eqnarray}
a_1&=&-\alpha b_1+\beta b_3, \quad a_3=\beta b_1+\alpha b_3, \nonumber
\end{eqnarray}
where
\begin{eqnarray}
\alpha&=& \frac{\gamma \theta n_2+\zeta}{\sqrt{2\zeta(\zeta+\gamma \theta n_2)}},\nonumber\\
\beta&=& \frac{\mathcal{J}}{2\sqrt{2\zeta(\zeta+\gamma \theta n_2)}},\nonumber
\end{eqnarray}
the effective Hamiltonian can be written in diagonalized form 
\begin{eqnarray}
H_{\rm eff} = \zeta (b_1^\dagger b_1-b_3^\dagger b_3),\nonumber
\end{eqnarray}
where we define
\begin{eqnarray}
\zeta = \sqrt{\left(\gamma \theta n_2\right)^2+\frac{\mathcal{J}^2}{4}}.\nonumber
\end{eqnarray}
Then, for the initial state $|\Psi(0)\rangle = |n_1,n_2,n_3\rangle$ the above Hamiltonian allows us to obtain the expectation value of population of sites 1 and 3
\begin{eqnarray}
\langle N_1\rangle &=& n_1+(n_3-n_1)\frac{\mathcal{J}^2}{4\zeta^2}\sin^2(\zeta t),\label{nk2}\\
\langle N_3\rangle &=& n_3-(n_3-n_1)\frac{\mathcal{J}^2}{4\zeta^2}\sin^2(\zeta t),\nonumber
\end{eqnarray}
which reduce to expressions given in \eqref{nk1} at $\theta = 0$. Figure \ref{dtheta} shows the comparison between the numerical simulation and the analytical result.
 From \eqref{nk2}, we observe that $\frac{\mathcal{J}^2}{4\zeta^2}\approx \left(\frac{J}{\gamma \theta n_2}\right)^2$ for $\gamma n_2$ sufficiently large, which shows that the oscillation amplitude of the atomic population of sites 1 and 3 decreases with order $O((\gamma \theta n_2)^{-2})$. Thus, the dynamics in the subsystem of sites 1 and 3 displays a faster localization of atomic populations under a small variation of angle $\theta$ in the vicinity of $\theta = 0$ in the strong interaction regime for the case $n_1+n_3\ll n_2$, as shown in Figure \eqref{dtheta}. This behavior is similar to that of a switching device \cite{PhysRevA.75.013608}, where the amplitude of the atomic flux between the source (well 1) and the drain (well 3) is controlled by the quantity of atoms initially present in the gate (well 2) under a small change in the direction of dipole polarization. In the first case (Figure \eqref{dtheta}-right-column), fine and smooth control of the tunnelling amplitude between source and drain is possible, which may be more desirable in situations involving the manipulation of quantum signals. On the other hand, the second case (Figure \eqref{dtheta}-left-column) shows an abrupt response to small variations in the dipole polarization direction, an essential feature that can be used to create more sensitive sensors, as we will see in the next section.
 
\section{\label{secV}Design of a quantum magnetic compass}
\begin{figure*}[!ht]
\center
\includegraphics[scale = 0.8]{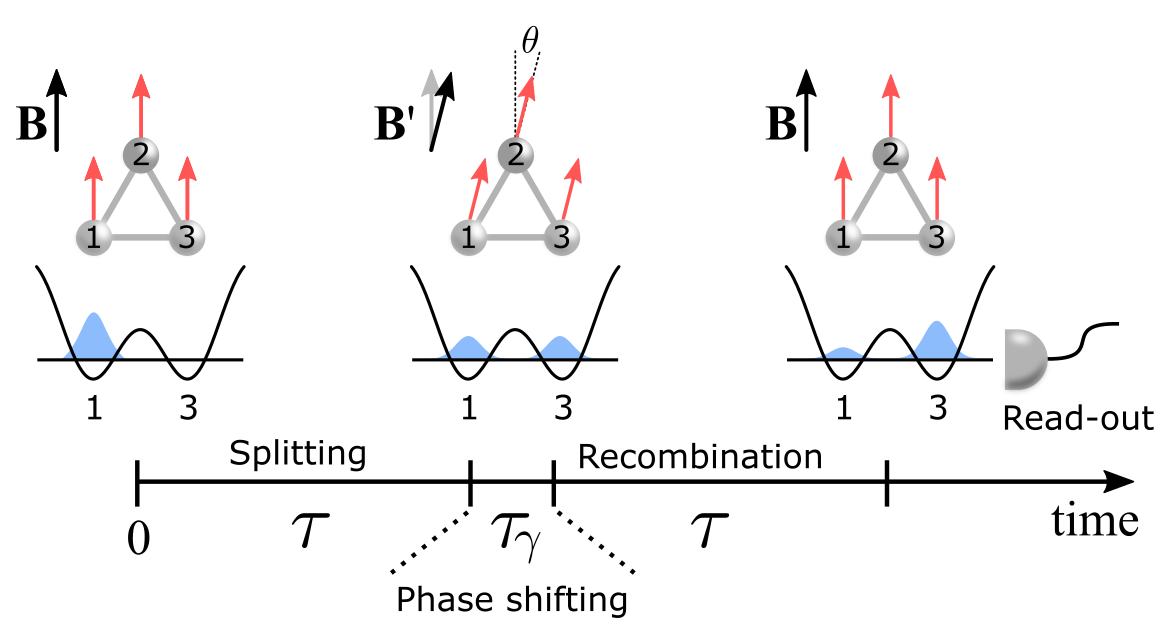}
\caption{Schematic representation of quantum magnetic compass illustrating the three-stage protocol, from the initial state preparation to the final measurement.}
\label{fig-schem}
\end{figure*}

\begin{figure}[!ht]
\center
\includegraphics[scale = 0.41]{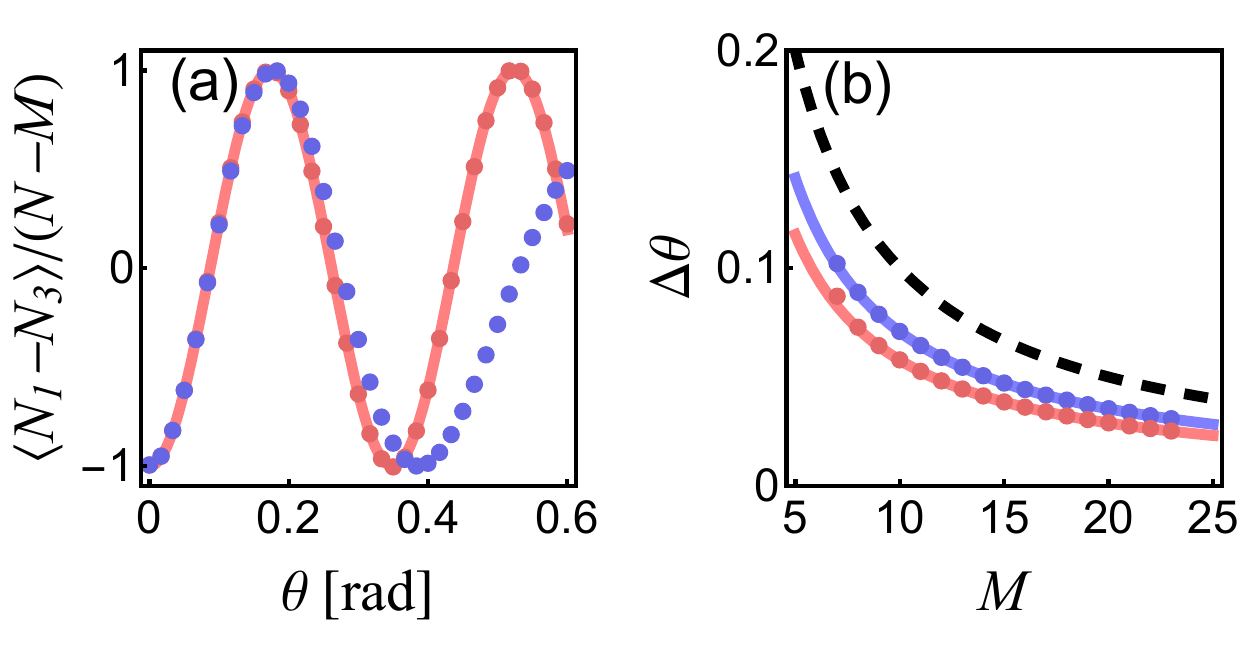}
\caption{(a) Imbalance population. Comparison between numerical simulation using the Hamiltonian \eqref{H} (red dotted) and \eqref{Hg} (blue dotted), and the formula \eqref{imb} (solid) using the initial state $|\Psi_0\rangle = |2,18,0\rangle$. (b) Phase error. The curves represent $\Delta \theta = 1/M$ (black dashed), $\Delta \theta = 1/(\sqrt{2}\,M)$ (blue solid) and $\Delta \theta = 1/(\sqrt{3}\,M)$ (red solid) for $\theta = 0.1$ rad. Numerical results using the Hamiltonian \eqref{H} are represented by dots. In all cases we consider $\nu/h = 0$.}
\label{fig-MZ}
\end{figure}

Based on the properties described previously, we now propose the design of a {\it quantum magnetic compass}, capable of detecting small changes in the direction of a uniform magnetic field acting on the triple-well circuit. For this purpose, we focus on the case $\phi = \pi/2$ and we use a Mach-Zehnder interferometry-like scheme \cite{Yurke1986interferometers}, following three steps (see Figure \ref{fig-schem}): ({\it splitting}) we start with a Fock state $|\Psi(0)\rangle =|N-M,M,0\rangle$, with $M\gg N-M$ and the dipoles polarized in the direction $\theta = 0$ by a magnetic field ${\bf B}$. We let the system evolve for a time interval $\tau = \pi/|2\mathcal{J}|$, necessary for the population of $N-M$ boson initially in well 1 to split between wells 1 and 3.  ({\it phase  shifting}) We then let a magnetic field ${\bf B}'$ with the same strength $|{\bf B}'| = |{\bf B}|=B$ but acting on the system in a slightly modified direction, $0<|\theta|\ll 1$, during the time interval $\tau_\gamma = 1/|2\gamma|$, resulting in a relative phase between the divided atomic populations. ({\it recombination}) Subsequently, we allow the atomic populations to recombine under the action of the field ${\bf B}$ in the direction $\theta = 0$ for a time interval $\tau$. Due to the relative phase, the recombination undergoes an interference process, resulting in a population imbalance between wells 1 and 3. Finally, the measurements of the populations in these wells signalize the change in direction of the magnetic field. The steps of this protocol can be represented by the state

\begin{eqnarray}
|\Psi(\theta)\rangle &=& \mathcal{U}(\tau,0)\mathcal{U}(\tau_\gamma,\theta)\mathcal{U}(\tau,0)|N-M,M,0\rangle,\nonumber
\end{eqnarray}
where $\mathcal{U}(t,\theta) = \exp(-it H)$ is the time evolution operator.
Using the effective Hamiltonian in the steps of protocol, we derive the imbalance population between sites 1 and 3:
\begin{eqnarray}
\langle N_1-N_3\rangle &=&\langle \Psi(\theta)|N_1-N_3|\Psi(\theta)\rangle\nonumber \\ &=& -(N-M)\cos(M\theta),\label{imb}
\end{eqnarray}
and its square
\begin{small}
\begin{eqnarray}
\langle (N_1-N_3)^2\rangle &&=\langle \Psi(\theta)|(N_1-N_3)^2|\Psi(\theta)\rangle \nonumber \\&&= (N-M)^2\cos^2(M\theta)+(N-M)\sin^2(M\theta).\nonumber
\end{eqnarray}
\end{small}
From \eqref{imb}, we observed that with a sufficient number of bosons in well 2, a small change in the angle $\theta$ results in a significant change in the imbalance population between wells 1 and 3, indicating that the phase $M\theta$ is super-resolution \cite{Mitchell2004}. 
This result is similar to those found in interferometers with a Heisenberg limit \cite{Dowling2002Rosetta}, where the sensitivity is characterized by the phase error $\Delta \theta = 1/M$. However, the estimation by the error propagation formula revels that  
\begin{eqnarray}
\Delta \theta=\frac{\Delta \langle N_1-N_3\rangle}{\left|\frac{d}{d\theta}\langle N_1-N_3\rangle\right|} = \frac{1}{M\sqrt{N-M}},\nonumber
\end{eqnarray}
where $\Delta \langle N_1 - N_3\rangle = \sqrt{\langle (N_1 - N_3)^2\rangle - \langle N_1 - N_3\rangle^2}$ is the standard deviation. This result indicates that the sensitivity of the quantum magnetic compass can beat the Heisenberg limit. Figure \eqref{fig-MZ}-a shows the comparison between expression \eqref{imb} and numerical simulation using the Hamiltonian \eqref{Hg} and \eqref{H}. We observe excellent agreement between the numerical and analytical results for $\theta < \pi/M\approx 0.17$ rad.  From then on, the results of the two Hamiltonians \eqref{Hg} and \eqref{H}  gradually disagree, as expected for higher values of $\theta$. Finally, Figure \eqref{fig-MZ}-b shows excellent agreement between the numerical simulation and the theoretical prediction of phase error when the difference between the populations of site 2 and the subsystem of sites 1 and 3 is sufficiently large.  

\section{\label{secVI}Discussions and overview}
We have investigated the dynamics of ultracold dipolar bosons confined in a triple-well circuit, subject to variations in the dipole polarization direction. Initially, we analyzed the role played by an external field in the band structure formation of the system's energy spectrum. We observed that, for sufficiently large values of the external field, the bands are energetically separated and that well-formed bands lead to harmonic dynamics of the atomic populations. For initial conditions in which the vast majority of bosons are in wells 1 and 3, the dynamics proved to be quite robust to gradual variations of the zenith angle $\phi$ at $\theta = 0$. In contrast, when the vast majority of atoms are in well 2, the dynamics remain robust to a wide variation in zenith angle $\phi$ at $\theta = 0$ in the absence of an external field. This preliminary study allowed us to identify the reliable configurations of the system in which the localization in the dynamics of the boson population is more sensitive to small variations of $\theta$ in the vicinity of $\theta=0$. In this scenario, we proposed the design of a quantum magnetic compass to detect small variation in the direction of uniform magnetic fields acting on the system. We found that this device shows a higher sensitivity than Heisenberg-limited devices. It is important to note that the proposed protocol does not require initialization in an entangled state in order to achieve enhanced sensitivity. This is possible because of the quadratic terms in the particle number operators of the Hamiltonian, resulting from the long-range DDI. Our study elucidates a complementary role played by dipole-dipole interaction in many-body physics and points to potential applications in the field of quantum sensing. In future research, we will extend our study to multiparameter quantum sensing involving systems with more modes in different geometries.


\newpage
\appendix

\section{Derivation of Hamiltonian}\label{appA}
We consider a system of dipolar boson confined into triple well potential trap $V_{\rm trap}({\bf r})$, where the centers of potential wells
\begin{eqnarray}
{\bf r}_1 = (-l/2,0,0),\quad {\bf r}_2 = (0,\sqrt{3}l/2,0),\quad {\bf r}_3 = (l/2,0,0),\nonumber 
\end{eqnarray}
are located at the vertex of equilateral triangle of side $l$ at the $xy$ plane. We assume that all dipoles are polarized along the direction of unit vector 
\begin{eqnarray}
{\bf d} = (\sin\phi \sin \theta,\sin\phi\cos\theta,\cos\phi),\nonumber
\end{eqnarray}
and the energies involved in the system are smaller than energy excitation to second Bloch band such that the on-site wave functions $\varphi_{i=1,2,3}({\bf r})=\varphi({\bf r}-{\bf r}_i)$ are determined by a Gaussian $\varphi$. Then, the second quantization theory leads to the extend Bose-Hubbard Hamiltonian \cite{Baranov2008, Jaksch1998}
\begin{eqnarray}
H_{\text{EBH}} &=& \frac{U_0}{2}\sum_{i=1}^3N_i(N_i-1)+\frac{1}{2}\sum_{i=j=1}^3 U_{ij}N_iN_j\nonumber \\ &-&J\sum_{i<j}^3(a_i^\dagger a_j+a_j^\dagger a_i).\nonumber
\end{eqnarray}
where $N_i=a_i^\dagger a_i$ and  $a_i\,(a_i^\dagger)$ are the particle number and annihilation (creation) operators of site $i$. The parameters of the system are given by
\begin{eqnarray}
U_0&=& U_{\rm s}+U_{\rm dip},\nonumber\\
U_{\rm s} &=& g\int d{\bf r} |\varphi_1({\bf r})|^4,\nonumber\\
U_{\rm dip} &=& \int d{\bf r} \, d{\bf r}'|\varphi_i({\bf r})|^2V_{\text{DDI}}({\bf r}-{\bf r}')|\varphi_i({\bf r}')|^2,\nonumber\\
U_{ij} &=& \int d{\bf r} \, d{\bf r}'|\varphi_i({\bf r})|^2V_{\text{DDI}}({\bf r}-{\bf r}')|\varphi_j({\bf r}')|^2,\nonumber\\
J&=& -\int d{\bf r}\varphi_1({\bf r})\left[-\frac{\hbar^2}{2m}\nabla^2+V_{\text{trap}}({\bf r})\right]\varphi_2({\bf r}),\nonumber
\end{eqnarray}
where the coupling $g= 4\pi \hbar^2 a /m$ characterizes the contact on-site interaction, $m$ is the mass of atom considered and the scattering length $a$ is controlled via Feshbach resonances. The on-site and inter-site interaction energies $U_{\rm dip}$ and $U_{ij}$ result from the DDI between dipolar boson, determined by the potential
\begin{eqnarray}
V_{\text{DDI}}({\bf r}) =\frac{\mu_0\mu^2}{4\pi}\frac{(1-3\cos^2\theta_P)}{|{\bf r}|^3},\nonumber 
\end{eqnarray}  
where $\mu_0$ is the vacuum magnetic permeability, $\mu$ is the permanent magnetic dipole moment of atom and $\theta_P$ is the angle between the direction of polarization ${\bf d}$ and the relative position ${\bf r}$ of the bosons. Using the DDI potential, the inter-site interaction energy can be written as
\begin{eqnarray}
U_{ij} = \widetilde{U}(1-3\cos^2\theta_{ij}),\nonumber
\end{eqnarray}
where $\widetilde{U}$ is the inter-site interaction energy when the angle $\theta_{ij}$ between the dipole polarization and the relative position between site $i$ and $j$ is $\theta_{ij} = \pi/2$. To calculate $\cos\theta_{ij}$, we consider the unit vector of relative position ${\bf  u}_{ij} = ({\bf r}_j-{\bf r}_i)/l$ of sites $i$ and $j$, from which we obtain
\begin{eqnarray}
\cos\theta_{13} &=& {\bf u}_{13}\cdot {\bf d} = \sin\phi\sin\theta,\nonumber\\
\cos\theta_{12} &=& {\bf u}_{12}\cdot {\bf d} = -\sin\phi\sin\left(\theta-\frac{2\pi}{3}\right),\nonumber\\
\cos\theta_{23} &=& {\bf u}_{23}\cdot {\bf d} = -\sin\phi\sin\left(\theta+\frac{2\pi}{3}\right).\nonumber
\end{eqnarray}
Next, using the identity
\begin{eqnarray}
\frac{1}{2}\sum_{i=1}^3N_i(N_i-1)&=&\frac{1}{2}N(N-1)-\sum_{i<j}^3N_iN_j,\nonumber
\end{eqnarray}
we redefine the Hamiltonian without the global constant
\begin{eqnarray}
H &=& H_{\text{EBH}}-\frac{U_0}{2}N(N-1)\nonumber \\&=&\sum_{i<j}^3(U_{ij}-U_0)N_iN_j-J\sum_{i<j}^3(a_i^\dagger a_j+a_j^\dagger a_i).\nonumber
\end{eqnarray}
We focus on the case where
\begin{eqnarray}
U_0 = \widetilde{U}.\nonumber
\end{eqnarray}
Then, defining
\begin{eqnarray}
U_\phi = -3\,\widetilde{U}\sin^2\phi ,\nonumber
\end{eqnarray}
the Hamiltonian can be conveniently rewritten as
\begin{widetext}
\begin{eqnarray}
H =U_\phi\left[ \sin^2\theta N_1 N_3+\frac{1}{4}\left[2+\cos(2\theta)\right]N_2(N_1+N_3)+\frac{\sqrt{3}}{4}\sin(2\theta)N_2(N_1-N_3)\right]
-J\sum_{i<j}^3(a_i^\dagger a_j+a_j^\dagger a_i).\nonumber
\end{eqnarray}
Now, using the identity
\begin{eqnarray}
N_2(N_1+N_3)&=&\frac{1}{4}\left[N^2-(N_1-N_2+N_3)^2\right],\nonumber
\end{eqnarray}
we obtain
\begin{eqnarray}
H &=&U_\phi\left[ \sin^2\theta N_1 N_3-\frac{1}{16}\left[2+\cos(2\theta)\right](N_1-N_2+N_3)^2+\frac{\sqrt{3}}{4}\sin(2\theta)N_2(N_1-N_3)\right]\nonumber \\
&-&J\sum_{i<j}^3(a_i^\dagger a_j+a_j^\dagger a_i)+\frac{U_\phi}{16}\left[2+\cos(2\theta)\right]N^2.\nonumber
\end{eqnarray}
Finally, defining
\begin{eqnarray}
U = -\frac{3}{16}U_\phi,\nonumber
\end{eqnarray}
and neglecting the global constant, we obtain
\begin{eqnarray}
H =\frac{U}{3}\left[2+\cos(2\theta)\right](N_1-N_2+N_3)^2-\frac{16}{3}U\left[ \sin^2\theta N_1 N_3+\frac{\sqrt{3}}{4}\sin(2\theta)N_2(N_1-N_3)\right]
-J\sum_{i<j}^3(a_i^\dagger a_j+a_j^\dagger a_i).\nonumber
\end{eqnarray}
\end{widetext}

\section{Experimental feasibility}\label{appB}
The triangular lattice can be generated by interfering three coplanar stand wave with wave length \break $\lambda =\frac{2\pi}{k}= 1.064$ $\mu$m, a vertical stand wave to control the aspect ratio of the potential wells and an optical dipole trap tightly focused with waist $w \sim 2\,\mu$m to confine the system of triple well into a plaquete, providing the potential
\begin{eqnarray}
V(x,y,z) = V_{\rm trap}(x,y,z) + V_{\rm ext}(x,y),\nonumber
\end{eqnarray}    
where
\begin{widetext}
\begin{eqnarray}
V_{\rm trap}(x,y,z)&=& V_0 \cos^2\left[\frac{k}{2} (\sqrt{3}x+y) \right] +V_0 \cos^2\left[\frac{k}{2} (-\sqrt{3}x+y) \right]+V_0 \cos^2\left(k y-\frac{\pi}{2}\right)
+ \frac{1}{2}m\omega^2z^2, \nonumber\\
V_{\rm ext}(x,y) &=& \frac{2V_1}{w^2}[x^2+(y-c-\Delta y)^2],\nonumber 
\end{eqnarray}
\end{widetext}
$V_0$ and $V_1$ are the potential depths, $m$ is the atomic mass, $l = \lambda/\sqrt{3}$ is the distance between the wells, $(0,c)$ with $c = l/(2\sqrt{3})$ is the location of centroid of the triple well and $\Delta y$ quantifies the displacement of the external field with respect to the centroid. 
The spherical potential can be obtained by trap frequency
\begin{eqnarray}
\omega&=&\sqrt{\frac{3k^2 V_0}{m} },\nonumber
\end{eqnarray}
and the potential of $i$-th well in the harmonic approximation is given by
\begin{eqnarray}
V^{(i)}(x,y,z) =\frac{1}{2}m\omega^2[(x-x_i)^2+(y-y_i)^2+z^2],\nonumber 
\end{eqnarray}
where the location of the wells at $xy$-plane are

\begin{eqnarray}
(x_1,y_1) &=& (-l/2,0), \qquad (x_2,y_2) = (0,\sqrt{3}l/2),\nonumber \\& &\qquad (x_3,y_3) = (l/2,0).\nonumber
\end{eqnarray}
The Wannier function of lowest Bloch band of site $i$ is represented by the Gaussian $\varphi_i({\bf r})=\varphi({\bf r}-{\bf r}_i)$, where
\begin{eqnarray}
\varphi({\bf r})= \left(\frac{2\eta}{\pi}\right)^2 e^{-\eta(x^2+y^2+z^2)}, \qquad \eta = \frac{m\omega}{2\hbar}.\nonumber
\end{eqnarray}
The contribution of external potential to Hamiltonian, $H\to H+H_{\rm ext}$, is given by
\begin{eqnarray}
H_{\rm ext} = \sum_{i=1}^3\nu_i N_i, \qquad \nu_i = \int d{\bf r}^3|\varphi_i({\bf r})|^2V_{\rm ext}({\bf r}).\nonumber
\end{eqnarray}
Using the above Gaussian, we obtain
\begin{eqnarray}
H_{\rm ext} = (\delta-\nu)N + \nu (N_1-N_2+N_3),\nonumber
\end{eqnarray}
where
\begin{eqnarray}
\delta &=& \frac{V_1}{w^2 \eta}\left[1+2\left(\frac{l^2}{4}+(c+\Delta y)^2\right)\eta\right],\nonumber\\
\nu &=& \frac{\sqrt{3} l\,\Delta y}{w^2}V_1.\nonumber
\end{eqnarray}
The first term of $H_{\rm ext}$ is a global constant and can be neglected.
\section{Hamiltonian parameters}\label{appC}
Defining the dimensionless parameter $q= l\sqrt{\eta}$, the dipolar interaction energy for dipoles polarized along $z$ direction can be calculated using the Fourier transform \cite{Lahaye_2009}
\begin{eqnarray}
\widetilde{U} = U_{ij}(q) = \frac{3\hbar^2 a_{\rm dd} }{ml^3}\left[\text{erf}(q)-\frac{2}{3\sqrt{\pi}}\,q\,e^{-q^2}\left(2\,q^2+3\right)\right],\nonumber
\end{eqnarray}
where
\begin{eqnarray}
a_{\rm dd}  =\frac{\mu_0\mu^2 m}{12\pi \hbar^2},\nonumber
\end{eqnarray}
is the dipolar length of atom. 
The above expression can be approximated using ordinary functions by \cite{Vadder1987}:
\begin{eqnarray}
\widetilde{U} = \frac{3\hbar^2 a_{\rm dd} }{ml^3}\left[\tanh(Cq+Dq^3)-\frac{2}{3\sqrt{\pi}}\,q\,e^{-q^2}\left(2\,q^2+3\right)\right],\nonumber
\end{eqnarray}
where $C = 167/148$ and $D=11/109$.
In spherical potential, $U_{\rm dip} = 0$ and the on-site interaction depends only contact interaction energy, which is given by
\begin{eqnarray}
U_0 = g\left(\frac{q}{l\sqrt{\pi}}\right)^{3}, \qquad g= \frac{4\pi \hbar^2 a}{m}.\nonumber
\end{eqnarray}
The condition $\widetilde{U}= U_0$, provides the value of scattering length $a=a_{\rm dd}\beta(q)$ where we defined 
\begin{eqnarray}
\beta(q)= \frac{3\sqrt{\pi}}{4q^3}\tanh(Cq+Dq^3)-\frac{e^{-q^2}}{2q^2}(2q^2+3).\nonumber 
\end{eqnarray}
The above function, has maximum value $\beta(q_0) \approx 0.217 $ at $q = q_0\approx 1.25$, which sets an upper limit for the scattering length. The Table \ref{tab} shows the experimental values used throughout the text to illustrate our results. 

\begin{table}[ht]
\centering
\caption{Experimental values calculated for dysprosium $^{164}$Dy atoms. We consider $\phi=\pi/2$, $\Delta y =l/2$ and $V_1 = V_0$, where $h$ is the Planck constant, $a_0$ is the Bohr radius, $E_r=h^2 /(2m\lambda^2)$ is the recoil energy and $m$ is the atomic mass.}
\begin{tabular}{lcc}
\hline
 &Parameters& Values \\
\hline
scattering length & $a$ & 11.1 $a_0$\\
wavelength &$\lambda$ & 1064 nm\\
dipolar length & $a_{\rm dd}$ & 131.97 $a_0$\\
potential depth & $V_0$ & 0.58 $E_r$\\
interaction energy & $\widetilde{U}/h$ & 5.38 Hz\\
hopping rate & $J/h$ & 1.55 Hz\\
trap frequency & $\omega/(2\pi)$ & 2.00 kHz\\
waist of gaussian beam & $w$ & 2 $\mu$m\\
energy offset & $\nu/h$ & 50.94 Hz\\
\hline
\label{tab}
\end{tabular}
\end{table}
\newpage

\nocite{*}

\bibliography{qcompass}

\end{document}